\documentclass[12pt]{article}
\def\be{\begin{equation}}
\def\ee{\end{equation}}
\def\bea{\begin{eqnarray}}
\def\eea{\end{eqnarray}}

\newtheorem{definition}{Definition}

\def\sup{\Sigma}
\def\suup{\sigma}
\def\eqq{\stackrel{\sup}{=}}
\def\mmm{{\cal W}}
\def\mm2{{\cal V}}
\def\varietat{{\cal V}}
\def\espaitemps{({\cal V},\g)}

\def\d{{\mbox d}}
\def\g{g}
\newcommand{\bm}[1]{\mbox{\boldmath $#1$}}

\def\R{{\rm I\!R}}

\def\Partp{\frac{\partial}{\partial \tilde{\varphi}}}
\def\Partz{\frac{\partial}{\partial \tilde{x}}}
\def\Partt{\frac{\partial}{\partial \tilde y}}
\def\Parsz{\frac{\partial}{\partial \zeta}}

\def\pparp{\partial/\partial \varphi}
\def\pparz{\partial/\partial x}
\def\ppartz{\partial/\partial \tilde{x}}
\def\ppartp{\partial/\partial \tilde{\varphi}}
\def\pparphi{\partial/\partial \phi}
\def\pparze{\partial/\partial \zeta}
\def\pparl{\partial/\partial \lambda}
\def\ppart{\partial/\partial \tilde y}

\def\lie{{\cal L}}
\def\cony{\vec \gamma}

\def\Journal#1#2#3#4#5#6{(#5) ``#6'' {#1} {\bf #2} #3#4}

\def\CQG{\em Class. Quantum Grav.}
\def\JPA{\em J. Phys. A: Math. Gen.}
\def\PRD{\em Phys. Rev. D }
\def\GRG{\em Gen. Rel. Grav.}

\def\JMP{\em J. Math. Phys.}

\def\CMP{\em Commun. Math. Phys.}

\def\PRL{\em Phys. Rev. Lett.}

\def\MPL{\em Mod. Phys. Lett.}

\def\AIHP{\em Ann. Inst. H. Poincar\'e}

\def\fin{\hfill \rule{2.5mm}{2.5mm}\\ \vspace{0mm}}
\def\finn{\hfill \rule{2.5mm}{2.5mm}}

\def\proof{\noindent{\em Proof.\/}\hspace{3mm}}

\newtheorem{lemma}{Lemma}[section]

\newtheorem{coro}{Corollary}[section]
\newtheorem{theorem}{Theorem}[section]

\def\bean{\begin{eqnarray*}}
\def\eean{\end{eqnarray*}}

\def\spI{(\mmm^I, g^I)}
\def\spE{(\mmm^E, g^E)}

\def\H{{\cal H}}

\begin{document}
\title{Symmetry-preserving matchings} 
\author{Ra\"ul Vera\\\\
School of Mathematical Sciences,
Queen Mary, University of London\\ 
Mile End Road                   
London E1 4NS, UK}
\date{July 2002}
\maketitle
\begin{abstract}
In the literature,
the matchings between spacetimes have been
most of the times implicitly assumed to preserve
some of the symmetries of the problem involved.
But no definition for this kind of matching was given
until recently.
Loosely speaking, the matching hypersurface is restricted
to be tangent to the orbits of a desired local group of symmetries
admitted
at both sides of the matching and thus admitted by the whole matched
spacetime.
This general
definition is shown to lead to conditions on the properties
of the preserved groups.
First, the algebraic type of the preserved group
must be kept at both sides of the matching hypersurface.
Secondly,
the orthogonal transivity of two-dimensional conformal
(in particular isometry) groups
is shown to be preserved (in a way made precise below)
on the matching hypersurface.
This result has in particular direct implications on the
studies of axially symmetric isolated bodies in equilibrium
in General Relativity,
by making up the first condition that determines
the suitability of convective interiors to
be matched to vacuum exteriors.
The definition and most of the
results presented in this paper
do not depend on the dimension of the manifolds involved
nor the signature of the metric,
and their applicability to
other situations and other higher dimensional theories 
is manifest.

\end{abstract}
\newpage
\section{Introduction}
\label{sec:intro}
Many practical problems of physical interest
involve idealised models constructed by joining
two regions, both admitting a certain symmetry\footnote{In what
follows only symmetries defined by local groups of
diffeomorphisms will be considered.
},
in such a way that the whole model admits
that symmetry globally.
These matchings
are then said to preserve the symmetry,
and have been extensively used, for instance, in
the description of astrophysical 
objects, the issue of the influence of cosmological
(global) dynamics on local systems and the description of hypersurface
layers, such as domain walls and `brane-world' scenarios.

Although the concept of symmetry--preserving
matching has an intuitive clear meaning, sometimes
its application has not fully properly
accounted for all the possibilities. 
For instance 
in some situations
the matching hypersurface has been prescribed `by hand',
making `hidden' implicit assumptions, such as
a preassigned meaning of the coordinates or the preservation of
geometrical properties, sometimes inferring
physical constraints to the model.
The problem here is that the lack of
generality prevents
conclusive results,
and hence the need for a general clear definition.
There have also been
some previous works \cite{ShLa,MASEuni}
where a general symmetry--preserving matching,
taking into account the orbits
of the preserved local groups at both sides of the matching hypersurface,
has been performed.
Nevertheless, no definition for symmetry--preserving matching had been given
in general form until recently \cite{raultesi}.

The purpose of this paper is twofold: In first place,
to motivate and present the definition of
symmetry--preserving matching
as stated in \cite{raultesi}. 
And secondly,
to show two immediate consequences on
the algebraic and geometric properties of the group
preserved by the matching.
These results
can lead to restrictions on the physical properties of the
model.
In fact, it will be shown here that some of the `hidden' assumptions mentioned
above can be derived from the junction conditions
as a consequence of the preservation of the symmetry.

Loosely speaking, the definition states that
the matching hypersurface is restricted
to be tangent to the orbits of a local group
wanted to be
admitted by the whole matched spacetime.
Its strict use
leads to more general parametrisations than usual,
which eventually manifests in more general ways of performing the matching,
some of them with clear physical differences.
The definition of symmetry--preserving matching
has already been proven very useful in obtaining
some general and conclusive results
\cite{MASEuni,raultesi,SEVEimp,MARCaxi,LRS-ES}.
In this sense,
in the study of the uniqueness problem of the exterior
solution given a known interior of an isolated axially symmetric
body in equilibrium in General Relativity
\cite{MASEuni},
the implicit use of the definition introduces two essential parameters
giving raise to inequivalent exteriors (if they exist).
One of these parameters represents, for example, the rotation
of the isolated body as seen by the stationary observer
at spatial infinity \cite{MASEuni}.
Another example where the introduction of parameters
by the matching procedure is crucial
has arisen in the study of the existence of locally cylindrically
symmetric static regions inside spatially homogeneous spacetimes
in four dimensions \cite{LRS-ES}.
As it will be explained below,
the existence of any static region of that kind in
a particular class of spatially homogeneous spacetimes is only possible
for non-zero values of one of the new parameters.
One could summarise the situation by saying that
the new parameters introduced
by the matching procedure are {\em locally} meaningless,
because they can always be absorbed in the coordinates,
but {\em globally} essential \cite{MASEuni,raultesi,nowahl,LRS-ES}. 

The definition has also geometric
consequences; first, the matching
hypersurface inherits the preserved symmetry and its algebraic type,
and as a result, the algebraic type of the preserved group will be
necessarily kept across the matching.
Secondly, the point-wise property that,
when satisfied in an open set,
ensures the orthogonal transitivity 
of any two dimensional conformal $G_2$ group
(including, of course, any $G_2$ group of isometries)
will also be kept
on the matching hypersurface
if the matching preserves that symmetry.
For simplicity,
and abusing the terminology,
I will refer to this by saying that
``the orthogonal transitivity is `preserved' on the matching
hypersurface'' in what follows.
In General Relativity, a very well known result
\cite{papapetrou,kundttrump,carter69,chuscomm}
can be used then
to ensure the extension of
the orthogonal transitivity
on a connected open set intersecting the matching 
hypersurface
when the $G_2$ is Abelian and
certain types of matter content are present.
These types include $\Lambda$-term type of matter, in particular vacuum,
and perfect fluids whose flow is orthogonal to the orbits
(if the $G_2$ acts on spacelike $S_2$ surfaces)
or tangent to the orbits (if the $G_2$ acts on timelike $T_2$ surfaces).

The `preservation'
of the orthogonal transitivity
on the matching hypersurface
will have a clear implication, for instance,
on the
studies of global models describing rotating axially symmetric
objects in equilibrium mentioned above.
These models are based on stationary and
axisymmetric spacetimes (admitting thus a $G_2$ on $T_2$)
consisting of the matching of a vacuum 
exterior region
to an 
interior region
across a timelike matching hypersurface that preserves
both the stationarity and the axial
symmetry. This hypersurface represents
the surface of the body at all times.
As a result of the existence of the axis of symmetry,
the $G_2$ on $T_2$ group on the exterior vacuum region
acts orthogonally transitively
because the Ricci tensor vanishes there
\cite{papapetrou}.
For convenience,
it has been usually assumed
an orthogonally transitive $G_2$ also in the interior.
This so-called {\em circularity condition\/} is equivalent to
the absence of convective motions in the interior,
and therefore restricts the physics of the astrophysical object.
Dropping the circularity condition in the interior problem
in order to deal with more general situations,
the result shown in this paper will imply
that the aforementioned property
coupled with the orthogonal transitivity holds
on the matching hypersurface 
for the $G_2$ on $T_2$
in the interior region.
In practical terms, this will
simplify
the structure of the matching conditions
for the more general 
problem.

Section \ref{sec:match} is devoted to a brief review
of the matching procedure in order to motivate and present
the definition of symmetry-preserving matching in Section \ref{sec:preserv},
where the preservation of the algebraic type is also shown.
The `preservation'
of the orthogonal transitivity for $G_2$ conformal groups
on the matching hypersurface
is addressed in Section \ref{sec:OT}.

Although the aim of this paper
has been originally focused on General Relativity,
the definition and results presented in this paper,
except Corollary \ref{teo:coro2},
neither depend on the dimension of the manifolds involved
nor on the signature of the metric.
Whenever the word ``spacetime'' is used in the definition
and the results,
it can perfectly be replaced by ``manifold with metric''.
The applicability to
other situations and other higher dimensional theories 
is manifest.

\section{Matching procedure}
\label{sec:match}
{}From the theoretical point of view,
given two $d+1$-dimensional
$C^3$ spacetimes $(\mmm^+,g^+)$ and $(\mmm^-,g^-)$
each of them with oriented boundary $\sup^+$ and $\sup^-$
\cite{HE}, respectively, such that
$\sup^+$ and $\sup^-$ are diffeomorphic, the whole matched spacetime
$\mm2$ is the disjoint union of both $\mmm^\pm$
with diffeomorphically related points in $\sup^\pm$ identified
and such that 
the junction conditions are satisfied
\cite{Darmois,LICH1,ISRA,CLDR,MASEhyper}.
In some cases, though,
certain junction conditions
will be relaxed in order to allow for distribution
layers, see below.
A brief review of the procedures involved in the practical setting of the
matching problem together with these junction conditions will be given
explicitly in what follows.

Let us consider the
practical problem \cite{FST} of the matching of two
given $C^3$ spacetimes $(\mm2^+,g^+)$ and $(\mm2^-,g^-)$.
First, we need
the $d$-dimensional hypersurfaces $\sup^+\subset \mm2^+$
and $\sup^-\subset \mm2^-$ which are going to
be identified.
As $\sup^\pm$ must be diffeomorphic to each other, they can also
be considered as diffeomorphic to an abstract
$d$-dimensional {\em oriented\/} $C^3$
manifold $\suup$ which can be appropriately
embedded both in $\mm2^+$ and $\mm2^-$.
Letting $\{\lambda^a\}$ (Latin indexes $a,b,\ldots=1,\ldots,d$)
be any local coordinate system on $\suup$ and
$\{x^{\pm\alpha}\}$ (Greek indexes $\alpha,\beta,\ldots=0,\ldots,d$) local
coordinates on $\mm2^\pm$, respectively, the two embeddings
are given by the $C^3$ maps
\bea
\label{eq:embeddings}
\Phi^\pm:\; \suup &\longrightarrow& \mm2^\pm\\
             \lambda^a &\mapsto& x^{\pm \alpha}
=\Phi^{\pm\alpha}(\lambda^a),
\nonumber
\eea
such that $\sup^\pm = \Phi^\pm(\suup)$. 
The diffeomorphism from $\sup^+$ to $\sup^-$
is trivially $\Phi^-\circ {\Phi^+}^{-1}$.
The hypersurfaces $\sup^\pm$
split locally each corresponding spacetime
$(\mm2^\pm,g^\pm)$ into two complementary parts.
These parts
are then the spacetimes with oriented boundary 
$({\mmm}_{1}^\pm,g^\pm,\sup^\pm)$
and $(\mmm_{2}^\pm,g^\pm,\sup^\pm)$
which are to be matched, and thus,
four possible different matchings are possible in principle,
although only two of them will be actually inequivalent \cite{FST}.
In order to simplify notations,
whatever part ${\mmm}_{1}^+$ or ${\mmm}_{2}^+$
is chosen to be matched with ${\mmm}_{1}^-$ or ${\mmm}_{2}^-$,
they will correspond to the theoretical ${\mmm}^+$ and ${\mmm}^-$.

Given the natural basis $\{\partial/\partial \lambda^a\}$
of the tangent bundle $T \suup$,
the rank-$d$ differential maps $\d\Phi^\pm$
apply $\{\partial/\partial \lambda^a\}$
into $d$ linearly independent
vector fields at $\sup^\pm$, denoted by
$\vec e^{\;\pm}_a|_{\sup^\pm}$,
defined only on the corresponding hypersurfaces $\sup^\pm$,
as follows
$$
\left.\d \Phi^\pm\right.\left(\left.\frac{\partial}{\partial
\lambda^a}\right.\right)
=\frac{\partial \Phi^{\pm\mu}}{\partial \lambda^a}
\left.\frac{\partial}{\partial x^{\pm\mu}}\right|_{\sup^\pm}
\equiv e^{\pm\mu}_a
\left.\frac{\partial}{\partial x^{\pm\mu}}\right|_{\sup^\pm}
=\vec e^{\;\pm}_a|_{\sup^\pm}.
$$
Using the pull-backs $\Phi^{\pm*}$,
the metrics $g^\pm$ can be mapped to $\suup$ providing
two symmetric 2-covariant tensor fields defined by
$\bar{g}^\pm\equiv
\Phi^{\pm *}(g^{\pm}|_{\sup^\pm})$.
These are the first fundamental forms of $\sup^\pm$, and in components
they read
\begin{equation}
\label{eq:pre}
\bar{g}^\pm_{ab}=
e^{\pm\mu}_a e^{\pm\nu}_b g^{\pm}_{\mu\nu}|_{\sup^\pm}.
\end{equation}
As shown in \cite{CLDR,MASEhyper}, the necessary an
sufficient
condition
such that there exists a {\em continuous\/} extension $g$ of the
metric to the whole manifold $\mm2$ and
such that $g|_{{\cal W}^+}=g^+$ and $g|_{{\cal W}^-}=g^-$
is that
\[
\bar{g}^+=\bar{g}^-.
\]
These relations were called the {\em preliminary junction
conditions}\footnote{Actually, the sufficient conditions
for the continuous extension of the metric
at points where the matching hypersurface is null
needs also the existence of
the two rigging vector fields with proper orientations as are defined later.
Their existence was erroneously stated in \cite{CLDR}.
I refer to \cite{signlong} for a detailed discussion and examples.}
in \cite{MASEhyper} and they have been traditionally used
since the work of Darmois \cite{Darmois}, Lichnerowicz \cite{LICH1}
and Israel \cite{ISRA}.
Once the above construction has been carried out,
after identifying the points $\Phi^+(p)=\Phi^-(p)\equiv q$
for all $p\in \suup$, 
the bases $\{\vec e^{\;+}_a|_q\}$ and $\{\vec e^{\;-}_a|_q\}$
can also be identified at every $q$.
I will denote simply by $\Sigma$ ($\equiv\sup^+=\sup^-$) the hypersuface in the
final matched manifold $\mm2$.

The one-forms normal to the hypersurfaces, denoted by $\bm n^\pm$,
are defined on $\sup^\pm$ up to a multiplicative non-zero factor
through the condition
$$
\bm n^\pm(\vec e^{\;\pm}_a)=0.
$$
In addition, $\bm n^+$ and $\bm n^-$ must
have the same norm in order to ensure their eventual proper
identification on the final matched spacetime $\mm2$.
The arbitrariness in the
sign in both $\bm n^+$ and $\bm n^-$
will account for their possible relative orientations
\cite{GOKA,FST}, see below.

In order to deal with general hypersurfaces (changing its causal
character from point to point),
one also needs two
transversal $C^2$ vector fields
$\vec\ell^{\;\pm}$ defined on $\sup^\pm$, the so-called rigging vectors
\cite{schouten}.
Obviously,
for the case of non-null hypersurfaces the normal vector is itself a rigging.
The riggings can be chosen by
\be
\bm n^\pm(\vec \ell^{\;\pm})=1
\label{eq:normirigg}
\ee
everywhere on $\sup^\pm$.
The vectors
$\{\vec \ell^{\;\pm},\vec e^{\;\pm}_a\}$ constitute a basis
of the tangent planes to $\mm2^\pm$ at any point on $\sup^\pm$.
The dual bases are given by $\{\bm n^\pm,\bm w^{\pm a}\}$
satisfying (dropping here $\pm$ everywhere)
\begin{equation}
  \label{eq:bases}
  \ell^\alpha {w^a}_\alpha=0,~~
  {w^a}_\alpha {e_b}^\alpha=\delta^a_b,~~
  n_\alpha {e_a}^\alpha=0,~~
  n_\alpha \ell^\alpha=1.
\end{equation}
Recalling that the preliminary conditions allowed us to identify
$\{\vec e^{\;+}_a\}$ with $\{\vec e^{\;-}_a\}$,
at this point there only remains to choose the riggings
such that $\{\vec \ell^{\;\pm},\vec e^{\;\pm}_a\}$ are both bases
with the same orientation and such that
\be
\ell^{+}_{\mu}\ell^{+\mu}\eqq \ell^{-}_{\mu}\ell^{-\mu}, \hspace{1cm}
\ell^{+}_{\mu}e^{+\mu}_{a}\eqq \ell^{-}_{\mu}e^{-\mu}_{a},
\label{eq:riggings}
\ee
where $\eqq$ means that both sides of the equality must be evaluated on
$\sup^{\pm}$ respectively.
Then, we can identify the whole $d+1$-dimensional
tangent spaces of $\mm2^\pm$ at $\sup$,
$\{\vec \ell^{\;+},\vec e^{\;+}_a\}=
\{\vec \ell^{\;-},\vec e^{\;-}_a\}\equiv\{\vec \ell,\vec e_a\}$,
as well as their respective dual co-bases
$\{\bm n^+,\bm w^{+ a}\}=\{\bm n^-,\bm w^{- a}\}
\equiv \{\bm n,\bm w^{a}\}$.
It must be taken into account that the first relation in (\ref{eq:riggings})
is implied by (\ref{eq:normirigg})
provided that the preliminary junction conditions and the second in
(\ref{eq:riggings}) hold.

The existence of a continuous $g$ allows for
the treatment of Einstein's equations in the distributional
sense \cite{ISRA,LICH2,taub1,CLDR,MASEhyper}.
Now,
the vanishing of the singular part of the Riemann tensor
distribution is equivalent to the equality
of the tensor fields defined 
by 
\begin{equation}
  \label{eq:H}
  \H^\pm_{ab}=
{e^\pm_a}^\mu
{e^\pm_b}^\nu \nabla^\pm_\mu \ell^\pm_\nu
\end{equation}
at both sides of $\sup$ \cite{MASEhyper}.
Of course, these latter junction conditions are omitted
in the studies focused on `surface' layer distributions,
such as topological defects, `brane-world' scenarios, and others.
For the case of non-null hypersurfaces,
$\H^{\pm}_{ab}$ coincide (up to a sign)
with the second fundamental forms
$K^{\pm}_{ab} =e^{\pm\mu}_{a}e^{\pm\nu}_{b}
\nabla^{\pm}_{\mu}n^{\pm}_{\nu}$
inherited by $\sup^{\pm}$ from
$\varietat^\pm$
\cite{Darmois,CLDR,MASEhyper} by choosing $\vec{\ell}= \pm \vec{n}$
(plus sign when $\vec{n}$ is spacelike and minus when timelike).
The junction conditions
\begin{equation}
  \label{eq:jc}
  \H^{+}_{ab}\eqq \H^{-}_{ab},
\end{equation}
do not depend on
the specific choice of the rigging \cite{MASEhyper}.

Notice that the choice of the same orientation for both
bases $\{\vec \ell^{\;\pm},\vec e^{\;\pm}_a\}$
is the same as choosing the riggings in such a way
that if $\vec\ell^{\;-}$ points from ${\cal W}^-$
outwards, then $\vec\ell^{\;+}$ points inwards onto ${\cal W}^+$,
or vice versa \cite{FST}.
The relative orientation of the rigging vectors
clearly translate to the relative orientations
of the normals through (\ref{eq:normirigg}).

\section{Definition of symmetry--preserving matching}
\label{sec:preserv}
In many practical problems of physical interest one looks
for a final whole spacetime having some symmetries and matched
across a hypersurface ``naturally'' defined by them.
This situation requires not only that both spacetimes to be
matched must contain these particular symmetries,
but also that the matching hypersurface inherits the
symmetry. This type of matching as well as the matching hypersurface
are usually said to {\em preserve the given symmetry}.
Illustrative simple
examples of matchings that preserve the symmetry are given by the
traditional spherically symmetric ones performed
across any time-dependent invariantly defined 2-sphere.
Of course, the idea behind all this is that we demand that
the matching hypersurface be tangent to the orbits of the
local symmetry group to be preserved.

Suppose that we are given two initial spacetimes
$(\varietat^\pm,\g^\pm)$ admitting the local groups of symmetries
$G_{n^{+}}$ and $G_{n^{-}}$ respectively.
Suppose also that we want the final matched spacetime
$(\varietat,\g)$  to have a local group of symmetries
$G_{m}$ which is a subgroup of both $G_{n^{\pm}}$,
so that $m\leq \min\{n^+,n^-\}$.
Here by local symmetries we mean not only isometries but more
general ones such as homothetic and conformal motions, etcetera
\cite{schouten,KRAM}.
The motivation for the definition as presented
here comes
from the consideration of symmetries that involve
the metric (including the conformal structure).
Despite this fact,
the definition does not depend on the kind of symmetry, and thus
symmetries that concern other objects, as for instance
collineations, projective symmetries, etc (see e.g. \cite{ghgensymm})
could also be considered.

Take any generator $\vec \xi$ of the subgroup $G_m$, such
that $\lie_{\vec\xi}\, \g$ has the corresponding form
(for instance, zero if $\vec\xi$ is a Killing vector field),
and assume that the restriction of $\vec\xi$ to
the hypersurface $\sup$
is {\em tangent\/} to $\sup$. This means that
\begin{equation}
\vec\xi~|_\sup=\xi^a \vec e_a|_\sup,
\label{eq:xie}
\end{equation}
where the $\xi^a$ are three
functions defined on $\sup\subset \mm2$.
Therefore, $\vec\xi$ provides 
in a natural way a unique vector field on $\suup$ \cite{schouten}
denoted by $\cony$ such that
\begin{equation}
  \label{eq:dmap}
  \d \Phi(\cony)=\vec\xi~|_{\sup}.
\end{equation}
A straightforward calculation shows then that \cite{schouten}
\begin{equation}
  \label{eq:first}
\lie_{\cony} \bar{\g}=
\Phi^* \left(\left.\lie_{\vec\xi}~ \g\right|_{\sup}\right).
\end{equation}
This result means that $\cony$ is a symmetry in $(\suup,\bar{\g})$
of the same (or more specialised) type as $\vec \xi$ is in
$\espaitemps$.
Notice that the fundamental property in all this
construction is the tangency of $\vec \xi$ to $\sup$.
In short, the symmetry defined by $\vec \xi$ in $\espaitemps$
is inherited by $\suup$ whenever $\vec \xi$ is tangent to its
image $\sup\subset\mm2$.
In fact, equation (\ref{eq:first}) allows us to show
the following:
\begin{lemma}
\label{teo:conf}
Let $\vec\xi^+$ and $\vec\xi^-$ be two conformal Killing
vector fields
acting on $(\mm2^+,g^+)$ and $(\mm2^-,g^-)$ respectively:
$\lie_{\vec\xi^\pm}g^\pm=\alpha^\pm g^\pm$ for
given, possibly zero, functions
$\alpha^\pm$.
If $(\mm2^+,g^+)$ and $(\mm2^-,g^-)$ are (preliminary) matched
across a matching hypersurface $\sup\equiv\sup^+=\sup^-$
diffeomorphic to $\suup$ by (\ref{eq:embeddings})
such that there is a vector field $\vec\gamma$ satisfying (\ref{eq:dmap})
for both $\vec\xi^+$ and $\vec\xi^-$, 
then $\alpha^+\eqq\alpha^-$.
\end{lemma}
\proof
Equation (\ref{eq:first}) for the ($+$) part reads
$$\lie_{\cony} \bar{\g}^+=
\Phi^{+*} \left(\left.\lie_{\vec\xi^+} \g^+\right|_{\sup^+} \right)=
\alpha^+|_{\sup^+}\Phi^{+*}(g^+|_{\sup^+})=\alpha^+|_{\sup^+}\bar{\g}^+,$$
and analogously for the ($-$) part.
The preliminary junction conditions
(\ref{eq:pre}) clearly imply $\alpha^+\eqq\alpha^-$ after
the identification $\sup\equiv\sup^+=\sup^-$.\fin

After the above discussion, it seems natural to give the following
definition of symmetry--preserving matching:

\begin{definition}
\label{def:mps}
Let $(\varietat,\g)$ be a spacetime arising from the matching 
of two oriented $C^3$ spacetimes $(\varietat^\pm,\g^\pm)$
admitting a $G_{n^{+}}$ and $G_{n^{-}}$ local group of symmetries,
respectively, and with respective boundaries $\sup^\pm$ given
by the embeddings (\ref{eq:embeddings}).
Then, $(\varietat,\g)$ preserves the symmetry
defined by the subgroup $G_m$ with $m\leq \min\{n^+,n^-\}$
when first, this group is admitted by both $(\varietat^\pm,\g^\pm)$,
and second, the differential maps $\d\Phi^\pm$ send
$m$ vector fields $\cony_A$ ($A=1\ldots m$)
on $\suup$ to the restrictions of the generators $\vec\xi^{\pm}_A$
of $G_m$ to $\sup^{\pm}$.
\end{definition}
{\it Remark:} It must be taken into account that if there is an intrinsically
distinguished generator of $G_m$
on $\varietat^+$ and $\varietat^-$,
then the symmetry--preserving matching
must ensure its identification
at $\sup$.
\\
\\
{}From the definition, and depending
on the kind of symmetry involved, it may follow that
$m$ has a maximum.
In the case of isometries, for instance,
this corresponds to the
case when the matching hypersurface $\Sigma$
is maximally symmetric, and hence $m\leq d(d+1)/2$ in this case,
being $d$ the dimension of $\sup$.

Of course, this definition is nothing but
the typical procedure used more or less
explicitly in the works on matchings that preserve symmetries.
In fact, in \cite{ShLa} Shaver and Lake already performed a matching
preserving the cylindrical symmetry taking into
account the orbits of the groups acting at both sides of
the junction (see also the less explicit procedures in \cite{flrwkas}).
Nevertheless, sometimes the conditions arising from
a strict use of the above definition leads
to parametrisations more general than the
usual ones in which the hypersurface is
assumed to preserve some further particular features of the
symmetries, such as orthogonal transitivity,
the meaning of the ignorable coordinates, and others
(see, for instance, \cite{MASEuni,raultesi,nowahl,LRS-ES}).
This definition has already been proven determinant in
recent works focused on General Relativity.
Thus, for instance, in \cite{MASEuni},
where the uniqueness of the
exterior solution given a known interior
in a matching of stationary axisymmetric spacetimes
is treated, it is shown how
the matching following Definition \ref{def:mps}
introduces two essential new parameters
giving raise to different exteriors, as it will be seen below.
Another example arises
in the study of the matching of $G_4$ on $S_3$ locally
rotationally symmetric spacetimes (LRS) with static cylindrically symmetric
ones preserving the cylindrical symmetry \cite{LRS-ES}.
In this case, if it were not by an essential parameter
introduced by the matching, no non-static
LRS model containing
a $G_3$ on $S_3$ subgroup of Bianchi types $V$ and $VII_h$
could be matched to any cylindrically symmetric static
spacetime.
In fact, in the same reference, the definition is also used
in order to show the existence of the axis of symmetry
at both sides of the matching hypersurface
if one of the halves is spatially homogeneous
and either that part or the other
represents a spatially compact and simply-connected
region \cite{LRS-ES}.

The generation of the parameterisations by the matching
procedure in these examples can be seen schematically as follows.
Let us have two 4-dimensional spacetimes
with Lorentzian metric $(\mm2^+,g^+)$ and $(\mm2^-,g^-)$
admitting a $G_2$ and a $G_3$ respectively,
both Abelian and including an axial symmetry,
to be matched preserving an Abelian $G_2$ containing the axial symmetry.
We know there exist coordinate systems on $(\mm2^+,g^+)$ and $(\mm2^-,g^-)$
where the Killing vector fields take the form
$\{\partial/\partial \varphi, \pparz\}$
and $\{\ppartp, \ppartz, \ppart\}$
respectively, where both $\partial/\partial \varphi$ and
$\partial/\partial \tilde\varphi$ generate an axial symmetry.
One starts by defining $\sup^+$.
Without loss of generality one can choose a coordinate $\phi$ on $\suup$
such that $\d\Phi^+(\pparphi)=\pparp$. Since the group is Abelian,
and after a suitable coordinate change on $\suup$ leaving
$\pparphi$ `unchanged', one can also choose a coordinate $\zeta$
on $\suup$ such that
$\d\Phi^+(\pparze)=\pparz$. Note that here we have used Lemma \ref{teo:algpre}
below, which ensures an Abelian $G_2$ on $\suup$.
The remaining coordinate $\lambda$ can
finally be chosen, without loss of generality,
such that $\pparl$ is sent to any vector field orthogonal
to both $\pparp$ and $\pparz$.
All this determines $\sup^+$ as
$\{\varphi=\phi,x=\zeta\}$ after a suitable choice of origin
of coordinates for $\phi$ and $\zeta$,
plus two arbitrary functions of $\lambda$
for the other two coordinates on $\mm2^+$.
Now, since the axial symmetry is uniquely defined
\cite{carter70,chuscomm,maseaxconf},
the vector $\pparphi$ must be sent to the axial generator in $\mm2^-$, i.e.
$\d\Phi^-(\pparphi)=\ppartp$.
On the other hand, $\pparze$ has to be mapped to {\em any}
Killing vector field that generates an Abelian $G_2$
together with $\pparp$. This implies that
\begin{equation}
\label{eq:gvect}
\d \Phi^-\left(\Parsz\right)=\left.a \Partp+b\Partz+c\Partt\right|_{\sup^-},
\end{equation}
where $a,b,c$ are constants.
The image of $\pparl$ through $\d \Phi^-$ will be a vector
orthogonal to the images of the other two.
As before,
$\sup^-$ is finally determined up to two arbitrary functions,
although its explicit expression in more involved than that of $\sup^+$.

Three parameters have been introduced by the matching in (\ref{eq:gvect}),
and although they can
eventually be restricted by the junction conditions,
they are free in principle.
The final remaining freedom introduced by these parameters
could be, in some cases where the metric $g^-$ is unknown,
absorbed by the metric functions after a coordinate change
of the kind\footnote{Clearly we do not want $b=c=0$ and hence
we can take $b\neq 0$ by interchanging $\tilde x$ and $\tilde y$
if necessary.}
\begin{equation}
  \label{eq:change}
 \tilde\varphi'=\tilde\varphi-\frac{a}{b}\tilde x,~~
 \tilde x'=\frac{1}{b}\tilde x,~~\tilde y'=\tilde y-\frac{c}{b}\tilde x.
\end{equation}
This transforms the vector at the right hand side of
(\ref{eq:gvect}) onto $\partial/\partial \tilde x'$
and leaves $\ppartp'=\ppartp$.
Nevertheless, as happens in \cite{MASEuni,LRS-ES},
the intrinsic meaning of some of the coordinates
$\{\tilde\varphi,\tilde x, \tilde y\}$ 
or the geometrical properties of the generators
invalidates the absorption of the parameters
after
the change (\ref{eq:change}) has been performed in most cases.

More explicitly, in \cite{MASEuni} both $(\mm2^\pm,g^\pm)$ 
are stationary and axisymmetric spacetimes.
Although in this case we do not necessarily
have a $G_3$ in $(\mm2^-,g^-)$,
the procedure
followed there can be described with the above
construction just by {\em not} considering $\partial/\partial \tilde y$
a Killing vector. We have to put then $c=0$ in (\ref{eq:gvect}).
The stationarity is accounted for
by taking $x$ and $\tilde x$
to be timelike coordinates.
In the asymptotically flat exterior $(\mm2^-,g^-)$,
there is a timelike coordinate, say $\tilde x$, with an intrinsic
meaning: it measures proper time at infinity.
Then, quoting from \cite{MASEuni},
if the interior describes a fluid with velocity vector
$\vec u=N(\pparz+ w \pparp)$, where $N$ and $w$ are two functions
that do not depend on $x$ nor $\varphi$, $\vec u$ on $\sup^-$
becomes
$\vec u|_{\sup}=N/b\left.\left[\ppartz + (wb-a)\ppartp\right]\right|_{\sup}$
by (\ref{eq:gvect}).
The proper angular velocity of the fluid on $\sup$ is then $(wb-a)$,
which depends on the parameters introduced by the matching $a,b$.
The change (\ref{eq:change}) (with $c=0$) could be used to absorb
the parameters {\em locally}, by redefining the metric functions,
but the {\em global} meaning of the coordinates clearly implies
a {\em global} meaning for the parameters $a,b$.

In \cite{LRS-ES}, $(\mm2^+,g^+)$ is taken to be a
$G_4$ on $S_3$ LRS spacetime. 
There is only a $G_2$ subgroup containing the axial
symmetry, hence Abelian \cite{alancil}, which will be
called $C_2$ in what follows.
On the other hand,
$(\mm2^-,g^-)$ corresponds to a static cylindrically symmetric
spacetime. The preserved symmetry is the Abelian $G_2$
(cylindrical symmetry), and thus
the above schematic construction makes sense 
by taking $x$ and $\tilde x$ to be spacelike coordinates
and $\tilde y$ a timelike coordinate.
In this case, and since there is no intrinsic meaning
for the coordinates {\em a priori} (and the metrics
are unknown at both sides), the parameters $a,b$
can indeed be absorbed after the change (\ref{eq:change})
into redefined metric functions so that we can put $a=0,b=1$
in (\ref{eq:gvect}) without loss of generality.
Nevertheless, the parameter $c$ has an important role to play:
In \cite{LRS-ES} it is also assumed that $(\mm2^-,g^-)$
admits an orthogonally transitive $G_2$  (see below), generated by,
say, $\ppartp$ and $\ppartz$.
The key point here is that
when the $G_4$ in $(\mm2^+,g^+)$
contains a $G_3$ subgroup of Bianchi types $V$ and $VII_h$
($h\neq 0$), the subgroup $C_2$ does not act
orthogonally transitively.
The preservation
of the orthogonal transitivity on $\sup$, which is going
to be shown in Section \ref{sec:OT},
implies in this case that if $c=0$ then, 
for non-static LRS $(\mm2^+,g^+)$,
the subgroup $C_2$ necessarily acts orthogonally transitively.
In short, not including $c$ in the matching procedure
would prevent non-static LRS spacetimes $(\mm2^+,g^+)$
with Bianchi types $V$ and $VII_h$
to be matched to the static $(\mm2^-,g^-)$.

The effect of the parameter $c$ is that
the $G_2$ generated by $\ppartp$ and $\ppartz+c\ppart$ for $c\neq 0$
is not orthogonally transitive.
As shown in \cite{LRS-ES}, having a general non-vanishing
$c$ allows LRS spacetimes $(\mm2^+,g^+)$ with
non-orthogonal transitive $C_2$ 
to be matched to the static $(\mm2^-,g^-)$ spacetimes
(which indeed contain an orthogonally transitive $G_2$).

\subsection{Preservation of the algebraic type}
\label{sec:pat}
The final matched manifold
$\espaitemps$ is at least $C^1$ (with $C^0$ metric),
but not necessarily $C^2$ across $\sup$.
Therefore the vector fields generating the $G_m$ group
at $\mm2$ are not differentiable across $\sup$
in general. As a consequence
the continuity of the algebraic type of the $G_m$ group
at both sides of $\sup$ cannot be ensured yet.
However, if $\sup$ preserves the symmetry
the algebraic type of $G_m$
must be also preserved across $\sup$.
This is obvious from the fact that the commutators
for $\cony_A$ are mapped by $\d\Phi^\pm$ to the commutators
of their respective $\vec\xi^{\pm}_A$ (\ref{eq:dmap}).
More explicitly,
taking $[\vec\xi^{\pm}_A,\vec\xi^{\pm}_B]=C^{\pm C}_{AB}\vec\xi^{\pm}_C$,
for any $C^1$ function defined on $\sup$, $f:\sup \to \R$,
we have \cite{schouten}
\begin{eqnarray*}
[\cony_A,\cony_B](\Phi^*f)&=&
\d \Phi\left([\cony_A,\cony_B]\right)(f)
=
\left.\left[\vec\xi_A,\vec\xi_B\right]\right|_{\sup}(f)
=\\
&=&C^C_{AB}\vec\xi_C|_{\sup}(f)=C^C_{AB}\cony_C (\Phi^*f),
\end{eqnarray*}
for both embeddings $\Phi^+$ and $\Phi^-$.
Hence $\suup$ inherits the algebraic type from both sides, i.e.
$C^{+ C}_{AB}\cony_C=[\cony_A,\cony_B]=C^{- C}_{AB}\cony_C$,
and thus the structure constants must coincide
at both sides for both sets of generators $\vec\xi^+_{A}$ and
$\vec\xi^{-}_{A}$ by construction.
Therefore, the definition above readily implies
that the group $G_m$ will have indeed the same algebraic type
at both sides of $\sup$.
This is summarised in the following lemma:
\begin{lemma}
\label{teo:algpre}
In a symmetry-preserving matching
the algebraic type of the preserved group $G_m$
is the same at both sides of the matching hypersurface.
If the $G_m$ corresponds to a conformal --not necessarily proper-- symmetry,
the matching hypersurface also inherits that symmetry (or a more
specialised one) and its algebraic type.\finn
\end{lemma}

This result may seem obvious, and it arises naturally
in the process of imposing the preliminary junction conditions
in a practical problem.
Nevertheless, it is essential in order to choose
the right coordinates on $\suup$ 
adapted to the symmetry given by $G_m$ and
its algebraic type. This has been more or less implicitly
assumed in the literature, but the explicit procedure
would follow the lines of the example concerning the preservation
of an Abelian $G_2$ above.

Furthermore, the setting of the matching hypersurface can be
simplified from the very beginning using Lemma \ref{teo:algpre},
see for instance \cite{LRS-ES}.
As an example, let us consider again the above situation,
but changing the algebraic type of the $G_3$
group in $(\mm2^-,g^-)$ from being Abelian to a more general one,
for instance a Bianchi $V$. The $G_3$ group contains then an Abelian $G_2$
subgroup, and again, this is the group we will preserve.
The coordinates $\{\tilde x^\alpha\}$
can be taken such that the generators are written as
$\{\ppartp,\ppartz,\vec v\}$ and $[\vec v,\ppartp]=\ppartp,
[\vec v,\ppartz]=\ppartz$. 
The same procedure follows until we get the analogous to equation
(\ref{eq:gvect}). We now have
\begin{equation}
  \label{eq:gvec2}
  \d \Phi^-\left(\Parsz\right)=\left.a \Partp+b\Partz+c\vec v\right|_{\sup^-}.
\end{equation}
We could carry on with all the matching procedure,
but Lemma \ref{teo:algpre} readily implies that we must have $c=0$,
because the vector at the right hand side in (\ref{eq:gvec2})
will have to commute with the image of $\pparphi$, this is $\ppartp$.

\section{Orthogonal transitivity of preserved conformal $G_2$ groups}
\label{sec:OT}
Given a $d+1$-dimensional spacetime admitting a two-dimensional
$G_2$ local group of symmetries
acting on non-null orbits,
the $G_2$ is said to be acting orthogonally transitively
(say, in an open set ${\cal U}$)
if there exists a family of $d-1$-surfaces
which are orthogonal to the orbits of the group.
Denoting by $\vec \xi,\vec \eta$ two independent
vector fields generating the group,
this happens iff
the two 4-forms defined by
$\bm \xi \wedge \bm \eta \wedge \d \bm \eta$ and
$\bm \xi \wedge \bm \eta \wedge \d \bm \xi$ vanish in  ${\cal U}$
(see e.g. \cite{schouten}).
In components, this is expressed as
$$
\xi_{[\alpha} \eta_{\beta} \nabla_{\mu} \eta_{\nu]} =
\xi_{[\alpha} \eta_{\beta} \nabla_{\mu} \xi_{\nu]} =0,
$$
where the square brackets stand for the usual antisymmetrisation.

This geometric property has important physical implications:
Global models for
astrophysical self-gravitating rotating bodies in equilibrium
in General Relativity
consist of stationary and axisymmetric
spacetimes,
thus admitting a $G_2$ local group of isometries
acting on timelike surfaces $T_2$, composed of two
main regions: an {\em interior} region $\spI$, that is to describe the
spatially compact and simply connected object,
and an {\em exterior} region $\spE$.
The two regions are matched across a timelike hypersurface $\sup$,
which describes the limiting surface of the body at all times.
The metric $g^I$ is taken to be a solution
of the Einstein field equations with matter,
whereas $g^E$ satisfies the equations for vacuum $R_{\alpha\beta}=0$.
Incidentally, if the model is to describe an isolated body, then
the exterior region is also taken to be asymptotically flat.

The existence of the axis of symmetry 
in the vacuum exterior $\spE$
implies that the $G_2$ on $T_2$ must act
orthogonally transitively there \cite{papapetrou}.
This very well known result
is based on the following
identities \cite{kundttrump,carter69,chuscomm,KRAM},
which are in fact valid for arbitrary dimension and signature,
for two Killing vector fields $\vec\xi$ and $\vec\eta$:
\begin{equation}
\nabla^\rho_{} \left( \eta^{}_{[\alpha} \xi^{}_\beta
\xi^{}_{\lambda;\rho]} \right)=
-\frac{1}{2}\xi^\rho_{} R^{}_{\rho[\lambda}\eta^{}_\alpha\xi^{}_{\beta]}+
\frac{1}{4} \left(
[\,\vec \eta,\vec \xi\,]^{}_{[\alpha}\nabla^{}_{\lambda}\xi^{}_{\beta]}+
\xi^{}_{[\alpha}\nabla^{}_{\lambda}[\,\vec \eta,\vec \xi\,]^{}_{\beta]}
\right),
\label{eq:ext}
\end{equation}
plus the $\xi\leftrightarrow \eta$ analogous.
The first term in the right hand side
vanishes if and only if the Ricci tensor
has an invariant 2-plane which is spanned
by the Killing vector fields at each point,
i.e.
\begin{equation}
\begin{array}{ccl}
&&R^{\alpha}_{\;\;\rho}\xi^{\rho}_{}=a_1(x^\beta)\,\xi^\alpha_{}+
b_1(x^\beta)\,\eta^\alpha_{},\\
&&R^{\alpha}_{\;\;\rho}\eta^{\rho}_{}=a_2(x^\beta)\,\xi^\alpha_{}+
b_2(x^\beta)\,\eta^\alpha_{},
\label{eq:ricciprop}
\end{array}
\end{equation}
where the functions $a$'s and $b$'s
need to satisfy some relations
to account for the symmetric character of $R_{\alpha\beta}$.
In the situation here,
the $G_2$ on $T_2$ groups acting in both
$\spI$ and $\spE$
must be Abelian because of the cyclic (axial) symmetry
\cite{carter69,jajora,alancil},
which leaves only the first term on the right hand side in (\ref{eq:ext}).
Now, equations (\ref{eq:ricciprop}) hold, for instance, when
the Ricci tensor (and hence the energy-momentum tensor)
is either proportional to the metric (so-called
$\Lambda$-term type, which includes vacuum) or of perfect-fluid
type and such that the fluid flow is orthogonal to the orbits
if the $G_2$ group acts on spacelike $S_2$ surfaces, or lies on the orbits
if the $G_2$ acts on timelike $T_2$ surfaces. This latter property corresponds
to the absence of convective motions
\cite{carter69}.

In the vacuum exterior region $\spE$ equations (\ref{eq:ricciprop})
hold trivially.
Denoting by $\{\vec\xi^E,\vec\eta^E\}$
the generators of the axial symmetry and stationarity in $\spE$ respectively,
$\bm \xi^E \wedge \bm \eta^E \wedge \d \bm \eta^E$ and
$\bm \xi^E \wedge \bm \eta^E \wedge \d \bm \xi^E$ are thus
constant\footnote{In four dimensions, the
codifferential of 4-forms vanish iff the 4-forms are constant.}
all over $\spE$, and equal to zero,
because $\vec\xi^E$ vanish at the axis of symmetry.

But regarding the interior region $\spI$
and denoting by $\{\vec\xi^I,\vec\eta^I\}$ the generators of the $G_2$
group there
which are eventually identified with $\{\vec\xi^E,\vec\eta^E\}$ on $\sup$,
although 
$\bm \xi^I \wedge \bm \eta^I \wedge \d \bm \eta^I$ and
$\bm \xi^I \wedge \bm \eta^I \wedge \d \bm \xi^I$
also vanish {\em at the axis},
they do not necessarily vanish everywhere in $\spI$,
because equations (\ref{eq:ricciprop})
do not hold {\em in principle}.
Hence, from equation (\ref{eq:ext}) it does not necessarily follow
that the $G_2$ on $\spI$
acts orthogonally transitively.
Nevertheless, in all the studies on global models describing rotating objects
in equilibrium it has been usually
assumed that the $G_2$ on $T_2$ group acts orthogonally
transitively {\em also} on the interior region
\cite{MERtesi,MASEuni,MarcERE,Marcexis}.
This is the so-called {\em circularity condition},
and, as mentioned above, implies the absence of convective motions
in fluids 
\cite{carter69}.

The matchings of spacetimes involved in the constructions
of such models 
have been
also always implicitly assumed to preserve
both the stationarity and the axial symmetry across
$\sup$. It is natural to ask then 
whether or not the symmetry--preserving matching
implies the preservation of
the orthogonal transitivity 
at the exterior across $\sup$
and hence leads to restrictions of the $G_2$ on $T_2$ in the interior.
Any result in this direction would turn the circularity
{\em condition} into a {\em consequence} of the
symmetry--preserving matching.

The following more general result ensures that the two 4-forms
$\bm \xi^I \wedge \bm \eta^I \wedge \d \bm \eta^I$ and
$\bm \xi^I \wedge \bm \eta^I \wedge \d \bm \xi^I$
defined above,
or equivalently the corresponding Hodge-dual (*) $d-3$-forms
(i.e. functions in four dimensions),
for the $G_2$ group at the interior must also vanish everywhere on $\sup$.

\begin{theorem}
Given a matching preserving the symmetry of a $G_2$
local conformal group --not necessarily proper-- as defined above, and choosing
$\{\vec\xi^+,\vec\eta^+\}$ and
$\{\vec\xi^-,\vec\eta^-\}$ as the sets of generators of
the $G_2$ groups at $(\mm2^+,g^+)$ and $(\mm2^-,g^-)$ respectively
such that 
$\d\Phi^{\pm}(\cony_1)=\vec\xi|_{\sup^\pm}^\pm$ and
$\d\Phi^{\pm}(\cony_2)=\vec\eta|_{\sup^\pm}^\pm$,
then
\begin{eqnarray}
  \label{eq:main}
 *(\bm \eta^+ \wedge \bm \xi^+ \wedge \d \bm \xi^+)&\eqq&
 *(\bm \eta^- \wedge \bm \xi^- \wedge \d \bm \xi^-),\\
  \label{eq:main2}
 *(\bm \xi^+ \wedge \bm \eta^+ \wedge \d \bm \eta^+)&\eqq&
 *(\bm \xi^- \wedge \bm \eta^- \wedge \d \bm \eta^-).
\end{eqnarray}
\label{teo:main}
\end{theorem}

\proof
Let us consider first $(\mm2^+,g^+)$.
The restriction of any 1-form field $\bm \xi^+$ to 
$\sup^+$
can be expressed in a co-basis $\{\bm n^+,\bm w^{a +}\}$
defined on $\sup^+$ and
dual to $\{\vec \ell^+,\vec e_a^+\}$ as
\begin{equation}
\bm\xi^+|_{\sup^+}
=\xi^+_a\bm w^{+ a}+\xi^+_\ell \bm n^+,
\label{eq:xi}
\end{equation}
where
$\xi^+_a=\bm \xi^+ ({\vec e^+_a})$ and $\xi^+_\ell=\bm \xi^+(\vec \ell^+)$.
Its exterior differential 2-form can be cast then as follows
\begin{equation}
  \label{eq:dxi}
  \d \bm \xi^+|_{\sup^+}
=A^+_{ab} ~\bm w^{+ a}\wedge\bm w^{b +}+
B^+_a ~\bm n^+\wedge \bm w^{+ a}
\end{equation}
with
\begin{eqnarray}
  \label{eq:AB2}
&&  A^+_{ab}=-A^+_{ba}={\vec e^+}_{[a}(\xi^{+}_{b]}),\nonumber\\
&&  B^+_a=2 \ell^{+ \alpha} {e^{+\beta}_a} \nabla^+_{[\alpha}\xi^+_{\beta]}=
\ell^{+ \alpha} {e^{+\beta}_a} \nabla^+_{\alpha}\xi^+_{\beta}
-{\vec e^+_a}(\xi^+_\ell) +
\xi^+_\alpha {e^{+\beta}_a} \nabla^+_\beta \ell^{+ \alpha}.
\end{eqnarray}
Another convenient expression for the latter is
\begin{equation}
B^+_a=\ell^{+ \alpha}{e^{+\beta}_a} \lie_{\vec\xi^+} g^+_{\alpha\beta}
-2{\vec e^+_a}(\xi^+_\ell) +
2 \xi^+_\alpha {e^{+\beta}_a} \nabla^+_\beta \ell^{+ \alpha}.
  \label{eq:AB}
\end{equation}
Taking the analogous expressions for another arbitrary 1-form $\bm \eta^+$,
the 4-form we look for reads then 
\begin{equation}
  \label{eq:4form1}
  \left.\bm \eta^+ \wedge \bm \xi^+
\wedge \d \bm \xi^+\right|_{\sup^+}=
\left.\left\{\eta^+_\ell \xi^+_c A^+_{ab}-
\eta^+_c\left(\xi^+_\ell A^+_{ab}-\xi^+_a B^+_b\right)\right\}\right.
\bm n^+\wedge \bm w^{+ a}\wedge \bm w^{+b}\wedge \bm w^{+c},
\end{equation}
on $\sup^+$. The analogous expressions follow when considering
the ($-$) counterpart.

We assume now that the preliminary junction conditions
hold, so that, following the construction as
explained in Section \ref{sec:match},
we have a whole matched spacetime $\espaitemps$ with
$\sup\equiv \sup^+=\sup^-$ splitting it into $\mmm^+(\subset \mm2^+)$
and $\mmm^-(\subset \mm2^-)$,
and such that 
the metric $g$ is continuous on the whole $\mm2$
and at least of class $C^2$ in both $\mmm^+$ and $\mmm^-$.
The tangent bases $\{\vec \ell^+,\vec e_a^+\}$ and
$\{\vec \ell^-,\vec e_a^-\}$
have been identified then to give $\{\vec \ell,\vec e_a\}$
at every $q\equiv \Phi^+(p)=\Phi^-(p)$,
as well as their respective cotangent bases have been identified
as $\{\bm n,\bm w^a\}$.
In other words, this means that
$\d\Phi^+(\partial/\partial \lambda^a)=
\d\Phi^-(\partial/\partial \lambda^a)$.

Taking now $\vec\xi^+$ and $\vec\xi^-$ to be two vector fields defined
on $\mm2^+$ and $\mm2^-$ respectively such that 
their restrictions
to $\sup^+$ and $\sup^-$ resp. are the images
through $\d\Phi^+$ and $\d\Phi^-$ of the same vector $\cony_1$ on $\suup$,
as follows from Definition \ref{def:mps},
the pair of vector fields $\vec\xi^+$ and $\vec\xi^-$
necessarily coincide on $\sup$.
And the same for
$\vec\eta^+$ and $\vec\eta^-$, which are images of the same $\cony_2$.
Since $g$ is continuous, then it obviously
follows that
$\xi^+_a=\xi^-_a\equiv \xi^{}_a$,
$\xi^+_\ell=\xi^-_\ell\equiv \xi^{}_\ell$,
i.e. $\bm \xi^+=\bm\xi^-$,
and analogously for $\eta^{}_a$ and $\eta^{}_\ell$,
at every point $q\in \sup$.
Defining $[f]\equiv f^+-f^-$ as
the difference of the values of the function $f$ on $\sup$ as
taken on $\sup^+$ and $\sup^-$, all this can be expressed as
$$
[\xi^{}_a]=0,~~ [\xi^{}_\ell]=0,~~ [\eta^{}_a]=0~~ [\eta^{}_\ell]=0.
$$
Any derivative along $\sup$ of a function $f$
satisfying $[f]=0$
also coincide as coming from either side, i.e. $[\vec e_a(f)]=0$,
and thus, in particular
\begin{equation}
  \label{eq:dstep}
  [\vec e_a(\xi^{}_b)]=0,~~ [\vec e_a(\xi^{}_\ell)]=0.
\end{equation}
As a result,
one gets
$$[A_{ab}]=
0.$$
Therefore, on $\sup$ one has
\begin{eqnarray}
\label{eq:diff4forms}
&&\left.\bm \eta^+ \wedge \bm \xi^+ \wedge \d \bm \xi^+\right|_\sup
-
\left.\bm \eta^- \wedge \bm \xi^- \wedge \d \bm \xi^-\right|_\sup=\\
&&\hspace{1cm}=\left[\eta_\ell \xi_c A_{ab}-
\eta_c\left(\xi_\ell A_{ab}-\xi_a B_b\right)\right]~
\bm n\wedge \bm w^{a}\wedge \bm w^{b}\wedge \bm w^{c}=\nonumber\\
&&\hspace{1cm}=\eta_c\xi_a [B_b]~
\bm n\wedge \bm w^{a}\wedge \bm w^{b}\wedge \bm w^{c}.\nonumber
\end{eqnarray}
This expression could have been obviously written as
$\bm \eta \wedge \bm \xi\wedge (\d \bm \xi^+-\d \bm \xi^-)$,
but I have preferred to keep the ($\pm$) signs on $\bm \xi$ and $\bm\eta$
throughout the proof for the sake of clarity. 
The coincidence of the two 4-forms across $\sup$ is then equivalent
to
\begin{equation}
\eta_{[a}\xi_b [B_{c]}]=0.
\label{eq:iff}
\end{equation}
{}From (\ref{eq:AB2}) and (\ref{eq:AB}) and their ($-$) counterpart, we have
\begin{equation}
  \label{eq:B}
  [B_a]=\ell^{\alpha}{e_a}^{\beta} [\nabla_{\alpha}\xi_{\beta}]
+\xi_\alpha [{e_a}^\beta\nabla_\beta \ell^{\alpha}]=
\ell^{\alpha}{e_a}^\beta [\lie_{\vec\xi} g_{\alpha\beta}]
+2 \xi_\alpha [{e_a}^\beta\nabla_\beta \ell^{\alpha}],
\end{equation}
where (\ref{eq:dstep}) has been used.
The first term in both expressions in
(\ref{eq:B}) is not zero in general, because
it contains derivatives of the components of $\vec\xi$
along $\vec\ell$, this is,
off $\sup$.
The second term in both expressions can be rewritten using (\ref{eq:xie})
as $\xi^b {e_b}^\alpha [{e_a}^{\beta} \nabla_\beta \ell_{\alpha}]$,
and thus, by the definition of the generalised second fundamental
form (\ref{eq:H}), (\ref{eq:B}) can be re-expressed as
\begin{equation}
  \label{eq:B2}
  [B_a]=
\ell^{\alpha}{e_a}^{\beta} [\nabla_{\alpha}\xi_{\beta}]
+\xi^b [\H_{ab}]=
\ell^{\alpha}{e_a}^\beta [\lie_{\vec\xi} g_{\alpha\beta}]
+2 \xi^b [\H_{ab}].
\end{equation}
Equation (\ref{eq:diff4forms}) reads then
\begin{equation}
  \label{eq:diff}
  \left.\bm \eta^+ \wedge \bm \xi^+ \wedge \d \bm \xi^+\right|_\sup
-
\left.\bm \eta^- \wedge \bm \xi^- \wedge \d \bm \xi^-\right|_\sup=
\eta_c\xi_a\left(\ell^{\alpha}{e_b}^\beta [\lie_{\vec\xi} g_{\alpha\beta}]
+2 \xi^d [\H_{bd}]\right)~
\bm n\wedge \bm w^{a}\wedge \bm w^{b}\wedge \bm w^{c}.
\end{equation}
So far we have only used the {\em preliminary junction conditions}
and the fact that $\vec\xi$ and $\vec\eta$ at both
sides are images of the same $\cony_1$ and $\cony_2$ respectively.
The second term at the right hand side in
(\ref{eq:diff})
clearly vanishes when imposing
the rest of the junction conditions (\ref{eq:jc}), $[\H_{ab}]=0$. 
On the other hand,
the first term at the right hand side in
(\ref{eq:diff}) vanishes if and only if
$\eta_{[a}\xi_b{e_{c]}}^\beta\ell^{\alpha}
[\lie_{\vec\xi} g_{\alpha\beta}]=0$, i.e.
\begin{equation}
\label{eq:lieg}
\ell^{\alpha} {e_{a}}^\beta[\lie_{\vec\xi} g_{\alpha\beta}]
=f_1~\xi_a+f_2~\eta_a,
\end{equation}
where the $f$'s are arbitrary functions defined on $\sup$.
In particular,
if $\vec\xi^\pm$ are conformal Killing vector fields,
Lemma \ref{teo:conf}, together with (\ref{eq:riggings}),
implies that
the left hand side of (\ref{eq:lieg}) vanishes,
and thus (\ref{eq:lieg}) is trivially satisfied for $f_1=f_2=0$.
In this case we have then
$\left.\bm \eta^+ \wedge \bm \xi^+ \wedge \d \bm \xi^+\right|_\sup
-
\left.\bm \eta^- \wedge \bm \xi^- \wedge \d \bm \xi^-\right|_\sup=0$.
It must be noted that (\ref{eq:lieg}), when the junction
conditions $[\H_{ab}]=0$ hold is, in fact, equivalent to
$\eta_{[a}\xi_b{e_{c]}}^\beta \ell^\alpha [\nabla_\alpha\xi_\beta]=0$
by (\ref{eq:B2}), this is
\begin{equation}
  \label{eq:jose1}
  \ell^{\alpha} {e_{a}}^\beta [\nabla_{\alpha}\xi_\beta]
=f_1~\xi_a+f_2~\eta_a.
\end{equation}
Since the metric $g$ is continuous and the orientation
has been preserved across $\sup$ we can take now
the Hodge-dual of this last expression.
This allows us to write
it as the equality of the corresponding Hodge-dual $d-3$-forms,
being $d+1$ the dimension of the manifolds.
Of course, the usefulness of this last step
is apparent in four dimensions.

The analogous equations and arguments follow for the difference
of the other 4-form $\bm \xi \wedge \bm \eta \wedge \d \bm \eta$
at both sides of $\sup$
by interchanging $\xi$ and $\eta$ in all the expressions above.
\fin

In particular, then, 
we have obtained the following:
\begin{coro}
\label{teo:coro}
Given a matching preserving a $G_2$ 
local conformal group --not necessarily proper--
as defined above and such that the $G_2$ acts orthogonally transitively
at one side of $\sup$, say at $(\mmm^+,g^+)$, then
\begin{equation}
  \label{eq:intcond}
  \begin{array}[c]{c}
  *(\bm \xi^I \wedge \bm \eta^I \wedge \d \bm \eta^I)\eqq0,\\

  *(\bm \xi^I \wedge \bm \eta^I \wedge \d \bm \xi^I)\eqq 0,
  \end{array}
\end{equation}
where $\{\vec \xi^-,\vec \eta^-\}$ are any two independent
generators of the $G_2$ group on $(\mmm^-,\g^-)$.\fin
\end{coro}

As a first remark,
let me stress again the fact
that in order to obtain (\ref{eq:main})-(\ref{eq:main2})
the assumptions on $\vec\xi$ and $\vec\eta$
being conformal Killing vector
fields as well as the junction conditions can be relaxed.
One only needs (\ref{eq:iff}), with (\ref{eq:B2})
(and their
$\xi\leftrightarrow\eta$ analogous), to be satisfied.
For this, as has been seen above, once all the junction conditions
are satisfied,
then only (\ref{eq:jose1}) (or, equivalently (\ref{eq:lieg}))
is needed.

{}From the converse point of view,
if (\ref{eq:jose1}) is satisfied
the necessary and
sufficient condition to obtain (\ref{eq:main})-(\ref{eq:main2})
is that $[\H_{bd}]$ has
an analogous geometric property as that of the Ricci tensor in
(\ref{eq:ricciprop}),
i.e.
\begin{equation}
\begin{array}{ccl}
\label{eq:necsuf}
&&\xi^d [{\H}_{bd}]=A_1~\xi_b+A_2~\eta_b,\\
&&\eta^d [{\H}_{bd}]=B_1~\xi_b+B_2~\eta_b,\\
\end{array}
\end{equation}
where, as above, the functions $A$'s and $B$'s defined
on $\sup$ satisfy certain relations
to account for the symmetric character of $[\H_{ab}]$.\footnote{$[\H_{ab}]$
is indeed symmetric, whereas $\H_{ab}$ is not in general \cite{MASEhyper}.}
This would be useful when relaxing Theorem \ref{teo:main}
from matchings of spacetimes to ``matchings'' allowing for
distributional parts on the Riemann tensor, as for instance,
in the description of topological defects and `brane-world' models.

As a second remark, note that
the result in Corollary \ref{teo:coro} only ensures 
the vanishing of the forms on a hypersurface ($\sup$).
The identities (\ref{eq:ext}) and its $\xi \leftrightarrow \eta$
analogous can be used then to extend this result
off $\sup$ in four dimensions. If the Ricci tensor (and thus the energy
momentum tensor in General Relativity)
satisfies (\ref{eq:ricciprop}) in $\mmm^-$
then (\ref{eq:ext}) (and the
$\xi\leftrightarrow \eta$ analogous) imply that
$*(\bm \eta^- \wedge \bm \xi^- \wedge \d \bm \xi^-)=
*(\bm \xi^- \wedge \bm \eta^- \wedge \d \bm \eta^-)= 0$
hold all over $\mmm^-$ and thus
the $G_2$ 
conformal local group acts orthogonally transitively there.
This can be summarised as follows:
\begin{coro}
\label{teo:coro2}
Given a matching in four dimensions preserving a $G_2$ 
local conformal group --not necessarily proper--
as defined above and such that the $G_2$ acts orthogonally transitively
at one side of $\sup$, say at $(\mmm^+,g^+)$,
if the Ricci tensor at the other side
$(\mmm^-,g^-)$ has an invariant 2-plane spanned by
two Killing vector fields generating the $G_2$ at each point,
then the $G_2$ acts orthogonally transitively in $(\mmm^-,g^-)$.\fin
\end{coro}

Corollary \ref{teo:coro} gives, in particular,
two necessary conditions for the existence
of the matching hypersurface.
If a spacetime admits a non-orthogonally
transitive $G_2$ such that (\ref{eq:intcond}) cannot be satisfied
anywhere, then it cannot be matched to a spacetime
admitting an orthogonally transitive $G_2$ whenever
the matching preserves those two $G_2$.
As an example, one can consider the specialised
van Stockum class of stationary cylindrically
symmetric dust solutions \cite{vanStockum,KRAM},
where the cylindrical symmetry is defined by a non-orthogonally
transitive $G_2$ on $S_2$ \cite{jajora}.
In this case it can be checked that
(\ref{eq:intcond}) is impossible, and thus
this solution cannot be matched to an orthogonally
transitive cylindrically symmetric spacetime while preserving
the cylindrical symmetry.
Nevertheless, it can
be matched to non-orthogonally transitive ones,
as was shown by van Stockum in \cite{vanStockum},
see also \cite{bonstock} and references therein.

Regarding the study of stationary and axisymmetric models,
the applicability of Corollary \ref{teo:coro}
is therefore immediate since it is the first
condition that determines the suitability
of convective interiors to be immersed in vacuum exteriors.
On the other hand, 
it also provides a first step in the generalisation
of the whole set of matching conditions
in the stationary and axisymmetric problem in General Relativity.
The whole set of matching conditions for the usual
stationary and axisymmetric
matchings with a vacuum exterior and a given orthogonally transitive
interior can be reorganised
in \cite{MERtesi,MASEuni} (a) conditions on the interior hypersurface,
(b) exterior matching hypersurface and
(c) boundary conditions for the exterior problem.
The conditions in (a) \cite{MERtesi,MASEuni}
correspond to an over-determined
system of ordinary differential equations given in this case by
the usual
direct consequence of the junction conditions (\ref{eq:jc})
on the discontinuities of the Einstein tensor $G_{\alpha\beta}$,
\begin{equation}
n^\alpha[G_{\alpha\beta}]=0.
\label{eq:israel}
\end{equation}
In the case of non-null matching hypersurfaces, these
are the so-called Israel conditions \cite{ISRA}.
In fact, and because of the symmetries
and the orthogonal transitivity involved in this case,
two of the four equations in (\ref{eq:israel})
are satisfied identically \cite{MASEuni}. Hence, two equations
arise at most from (\ref{eq:israel}), 
whose compatibility 
is then necessary for the existence of $\sup^I$.
In this case, and for perfect fluid interiors,
the Israel conditions translate into the intuitive vanishing of the
pressure at $\sup^I$.
The equation $p\eqq0$ defines then the
matching hypersurface as seen from the interior $\sup^I$
in an implicit manner.
In some occasions the equations
(\ref{eq:israel})
are satisfied identically, as for example the case of dust.
In these cases $\sup^I$ is not determined and one could match, in principle,
across any timelike hypersurface preserving the
symmetry.

When $\sup^I$ is uniquely defined,
conditions (b) (see \cite{MERtesi,MASEuni}) determine $\sup$ as seen
from the exterior, i.e. $\sup^E$.
The rest of the matching conditions (c) provide
the over-determined
boundary conditions for the elliptic vacuum exterior problem
by giving the values of the Ernst potential
up to an additive constant
and its normal derivatives on $\sup^E$ \cite{MASEuni}.

If the circularity condition is dropped, and
because neither $\vec\xi^I$ nor $\vec\eta^I$ vanish all over the
matching hypersurface,
the two necessary conditions (\ref{eq:intcond})
constitute two more equations defining $\sup^I$.
Furthermore, and because now there are more non-zero components
of the Einstein tensor, equations (\ref{eq:israel})
will result on more relations.
But it still remains to be checked whether or not
the Israel conditions (\ref{eq:israel}) plus that in (\ref{eq:intcond})
account for all the 
matching conditions determining the existence
and eventual definition of $\sup^-$.
The existence of dust solutions
admitting non-orthogonally transitive $G_2$ groups of isometries
(see e.g. \cite{vanStockum,bonstock,king74,KRAM})
ensures that the Israel conditions (\ref{eq:israel})
do not imply the new conditions (\ref{eq:intcond})
in general.

\section{Conclusions}
After motivating and presenting
the definition of symmetry--preserving matching
as introduced in \cite{raultesi},
this paper has dealt with the first consequences
such definition has on the preserved group.
A usual property of the differential maps
has been implemented here to see how
the algebraic type of the preserved group must be kept 
at both sides of the matching hypersurface.

It has also been shown the `preservation'
of the orthogonal transitivity of conformal $G_2$ groups
on the matching hypersurface.
The implications of this result on the generalisation
of the studies of stationary and axisymmetric
models of isolated bodies in General Relativity
to allow for convective interiors
have been discussed.
The next step at this point should be to address the problem of
the whole set of matching conditions when the interior admits
a general Abelian $G_2$. 
Since the only new property, the non-orthogonal
transitivity in the interior, is driven by two
functions that have to vanish on $\sup^I$ by Corollary \ref{teo:coro},
it may seem that the only new terms to be added in the matching conditions
would come from the normal derivatives of the two functions
$* (\bm \xi^I \wedge \bm \eta^I \wedge \d \bm \eta^I)$ and
$* (\bm \xi^I \wedge \bm \eta^I \wedge \d \bm \xi^I)$.
These new terms would
appear by modifying the expressions for the values of the normal
derivatives of the exterior Ernst potential on $\sup$
as obtained in the usual non-convective case \cite{MASEuni}.
The study of the whole general set of matching conditions
for a general Abelian $G_2$ in the interior
is currently under investigation.
Obtaining an analogous structure of equations for the matching
in the general case as in the non-convective case
would be very useful,
because all the results and studies on uniqueness \cite{MASEuni}
and existence \cite{MarcERE,Marcexis}
of the exterior vacuum problem could be implemented
in a straightforward manner to the general case without
the need of the circularity condition.

\section*{Acknowledgements}
I am grateful to Jos\'e Senovilla and Marc Mars
for several very careful readings of this manuscript
and for their many fruitful ideas, discussions and
valuable criticisms.
I would also like to thank Malcolm MacCallum and Filipe Mena
for pointing out some references and the careful reading
of the manuscript.
I also thank EPSRC for grant MTH 03 R AJC6.



\end{document}